\documentstyle[psfig,12pt]{article}
\def\TL{\hfil$\displaystyle{##}$}
\def\TR{$\displaystyle{{}##}$\hfil}
\def\TC{\hfil$\displaystyle{##}$\hfil}
\def\TT{\hbox{##}}
\def\seqalign#1#2{\vcenter{\openup1\jot
  \halign{\strut #1\cr #2 \cr}}}

\def\fixit#1{}

\def\mop#1{\mathop{\rm #1}\nolimits}

\def\overleftrightarrow#1{\vbox{\ialign{##\crcr
     $\leftrightarrow$\crcr\noalign{\kern-0pt\nointerlineskip}
     $\hfil\displaystyle{#1}\hfil$\crcr}}}

\def\lsim{\mathrel{\mathstrut\smash{\ooalign{\raise2.5pt\hbox{$<$}\cr\lower2.5pt\hbox{$\sim$}}}}}
\def\gsim{\mathrel{\mathstrut\smash{\ooalign{\raise2.5pt\hbox{$>$}\cr\lower2.5pt\hbox{$\sim$}}}}}
\def\sqr#1#2{{\vcenter{\vbox{\hrule height.#2pt
         \hbox{\vrule width.#2pt height#1pt \kern#1pt
            \vrule width.#2pt}
         \hrule height.#2pt}}}}
\def\square{\mathop{\mathchoice\sqr56\sqr56\sqr{3.75}4\sqr34\,}\nolimits}
\def\href#1#2{#2}  

\def\lbldef#1#2{\expandafter\gdef\csname #1\endcsname {#2}}
\def\eqn#1#2{\lbldef{#1}{(\ref{#1})}%
\begin{equation} #2 \label{#1} \end{equation}}
\def\eqalign#1{\vcenter{\openup1\jot
    \halign{\strut\span\TL & \span\TR\cr #1 \cr
   }}}
\def\eno#1{(\ref{#1})}
\def\Re{\mop{Re}}
\def\Im{\mop{Im}}

\begin{document}
\baselineskip=15.5pt
\pagestyle{plain}
\setcounter{page}{1}
%\renewcommand{\thefootnote}{\fnsymbol{footnote}}
%--------+---------+---------+---------+---------+---------+---------+
%Title page

\begin{titlepage}

\begin{flushright}
PUPT-1966 \\
hep-th/0011127
\end{flushright}
\vfil

\begin{center}
{\huge The evolution of unstable black holes}
\vskip0.5cm
{\huge in anti-de Sitter space}
\end{center}

\vfil
\begin{center}
{\large S. S. Gubser and I. Mitra}
\end{center}

$$\seqalign{\span\TL & \span\TT}{
& Joseph Henry Laboratories, Princeton University, Princeton,
NJ 08544
}$$
\vfil

\begin{center}
{\large Abstract}
\end{center}

\noindent 
 We examine the thermodynamic stability of large black holes in
four-dimensional anti-de Sitter space, and we demonstrate numerically
that black holes which lack local thermodynamic stability often also
lack stability against small perturbations.  This shows that no-hair
theorems do not apply in anti-de Sitter space.  A heuristic argument,
based on thermodynamics only, suggests that if there are any
violations of Cosmic Censorship in the evolution of unstable black
holes in anti-de Sitter space, they are beyond the reach of a
perturbative analysis.

\vfil
\begin{flushleft}
November 2000
\end{flushleft}
\end{titlepage}
\newpage
%--------+---------+---------+---------+---------+---------+---------+
%Body
\section{Introduction}
\label{Introduction}

The Gregory-Laflamme instability \cite{glOne} is a classical
instability of black brane solutions in which the mass tends to clump
together non-uniformly.  The intuitive explanation for this
instability is that the entropy of an array of black holes is higher
for a given mass than the entropy of the uniform black brane.  The
intuitive explanation leaves something to be desired, since it applies
equally to near-extremal D$p$-branes: scaling arguments establish that
a sparse array of large black holes threaded by an extremal D$p$-brane
will be entropically favored over a uniform non-extremal D$p$-brane;
however it is not expected that near-extremal D$p$-branes exhibit the
type of instability found in \cite{glOne}.  It was checked in
\cite{glTwo} that a D$p$-brane which is far from extremality (that is,
one whose tension is many times the extremal tension) does have an
instability.  It was also shown that the instability persists for
charged black strings in five dimensions fairly close to
extremality.\footnote{The charged black string studied in \cite{glTwo}
happens to be thermodynamically unstable all the way down to
extremality: the specific heat is negative.  Thus \eno{Conjecture}
would lead us to believe that this non-extremal black string is always
unstable.  The extremal solution should be stable since it can be
embedded in a supersymmetric theory as a BPS object.}  Less is known
about the case of near-extremal D3-branes, M2-branes, and M5-branes,
but one may take the absence of tachyons in the extensive AdS-glueball
calculations (\cite{witHolTwo,Ooguri} and subsequent works---see
\cite{MAGOO} for a review) as provisional evidence that these
near-extremal branes are (locally) stable.\footnote{More properly, we
should say that the near-extremal black brane solutions with many
units of D3-brane, M2-brane, or M5-brane charge appear to be stable.
A single brane has Planck scale curvatures near the horizon, so
classical two-derivative gravity does not provide a reliable
description.  We will concern ourselves exclusively with solutions
which have a discrete parameter (M2-brane charge, for the most part)
which can be dialed to infinity to suppress all corrections to
classical gravity.}

In the AdS/CFT correspondence \cite{juanAdS,gkPol,witHolOne}, one
might at first think that the existence of a unitary field theory dual
forbids an instability.  But suppose we are at finite temperature, and
that there is a thermodynamic instability in the field theory---like
the onset of a phase transition.  Then it is quite natural for some
fluctuation mode (or modes) to grow exponentially in time, at least in
a linearized analysis, as one nucleates the new phase.  Exciting an
unstable mode is a change in the state of the field theory, not its
lagrangian; thus according to AdS/CFT there should be a {\it
normalizable} mode in AdS which likewise grows exponentially with time
\cite{bkl}.  This might be referred to as a ``boundary tachyon,'' or a
``tachyonic glueball,'' since in the gauge theory it corresponds to
some bound state with negative mass-squared.  We will prefer the term
``dynamical instability,'' which is meant to convey that there is an
instability in the Lorentzian time evolution of the black brane, in
both its supergravity and dual field theory descriptions.

To sum up, the existence of a field theory dual makes plausible the
following adaptation of the entropic justification for the
Gregory-Laflamme instability:
  \eqn{Conjecture}{\eqalign{
   &\hbox{\it For a black brane solution to be free of dynamical instabilities,
    it is necessary}  \cr
   &\hbox{\it and sufficient for it to be locally thermodynamically stable.}
  }}
 Here, local thermodynamic stability is defined as having an entropy
which is concave down as a function of the mass and the conserved
charges.  This criterion was first used in a black brane context in
\cite{gspin}, where it was found that spinning D3-branes could be made
locally thermodynamically unstable if the ratio of the spin to the
entropy was high enough.  Further work in this direction, relevant to
the current paper, has appeared in \cite{cgOne,cgTwo,Cai,HarmarkOne}.
For a somewhat complementary point of view on the nature of the
unstable solutions, see \cite{cejmOne,cejmTwo}.

The conjecture \Conjecture\ is meant to be a local version of the
argument about whether the array of black holes or the black brane has
higher entropy; however it seems on more precarious ground since one
may not be able to write down a non-uniform stationary solution that
competes with the black brane entropically.  Nonetheless, it was shown
in \cite{gMitra} that \Conjecture\ predicts with good accuracy the
value of the charge where the four-dimensional anti-de Sitter
Reissner-Nordstrom solution ($AdS_4$-RN) develops an instability.

The aim of this paper is to give a fuller exposition of the
calculations in \cite{gMitra}, to present a more complete picture of
the results of the numerics, and to explore via thermodynamic
arguments the likely paths for time-evolution of the unstable
solutions.  Section~\ref{Solutions} contains a summary of the
$AdS_4$-RN solution and some generalizations of it in ${\cal N}=8$
gauged supergravity and in higher dimensions.  Section~\ref{Thermo}
discusses the thermodynamic instability which occurs for large charge.
In section~\ref{Instability}, a linear perturbation analysis is
carried out around the $AdS_4$-RN solution.  In the large black hole
limit, a dynamical instability appears when local thermodynamic
stability is lost.  The existence of a dynamical instability was the
main result of \cite{gMitra}.  It disproves the claim of
\cite{Hawking,HawkingStrings} that charged black holes in AdS are
classically stable.  As we explain in section~\ref{Instability}, the
instability persists some ways away from the large black hole limit,
providing the first proven example of a black hole with a compact
horizon and a pointlike singularity which exhibits a dynamical
Gregory-Laflamme instability.\footnote{Here we are referring to the
existence of a local instability visible in a classical analysis.  It
has been observed \cite{bdhm} that the AdS-Schwarzschild solution
times a sphere can have a lower entropy than a Schwarzschild black
hole of the same mass which is localized on the sphere.  This
demonstrates global but not local instability, and suggests the
possibility of tunneling from one configuration to the other.}  Such
solutions are interesting from the point of view of Cosmic Censorship,
and we discuss the possibility of forming a naked singularity, or at
least regions of arbitrarily large curvatures.  Our main result here
is that adiabatic evolution toward maximum entropy does not lead to
solutions which arise from making the mass smaller than some
appropriate combination of the charges.  Because entropic arguments
appear to give good information not only on the existence of dynamical
instabilities but also on the direction they point, it is reasonable
to predict from our results that no perturbative analysis of a smooth
black hole in AdS will demonstrate a violation of Cosmic Censorship.

The unstable mode of the $AdS_4$-RN solution does not involve
fluctuations of metric at linear order.  Rather, it involves the gauge
fields and scalars of ${\cal N}=8$ gauged supergravity.  Because the
metric is not fluctuating, it may seem odd to describe the process as
a Gregory-Laflamme instability.  But we claim that the instability we
see is in the same ``universality class'' as instabilities where the
horizon does fluctuate: to be more precise, if the charges of the
black hole are made slightly unequal, then generically the instability
will involve the metric.  In fact, the metric does fluctuate in the
equal charge case as well---only at a subleading order that is beyond
the scope of our linearized perturbation analysis.  We would in fact
make the case that any dynamical instability of a black hole which
leads to non-uniformities in charge or mass densities should be
considered in the same category as the Gregory-Laflamme instability of
uncharged black branes.

We emphasize that this paper is concerned with the relation between
local thermodynamic stability of stationary solutions and the
stability of their classical evolution in Lorentzian time.  It is
known
\cite{HawkingPage,Perry,PrestidgeOne,PrestidgeTwo,PecaOne,PecaTwo} 
that black
holes which are thermodynamically unstable have an unstable mode in
the Euclidean time formalism.  For spherically symmetric black holes
this mode is an $s$-wave.  The interpretation is that, for instance,
an AdS-Schwarzschild black hole in contact with a thermal bath of
radiation will not equilibrate with the bath if the specific heat of
the black hole is negative.  This beautiful story does not fall under
the rubric of problems we are considering, because the processes by
which equilibration takes place in Lorentzian time include Hawking
radiation, which is non-classical.  Rather, we are contemplating black
holes or branes in isolation from other matter, in a classical limit
where Hawking radiation is suppressed, and inquiring whether a
stationary, uniform black object wants to stay uniform or get lumpy as
Lorentzian time passes.  It is less clear that there should be any
relation between this dynamical question and local thermodynamic
stability: for instance, a Schwarzschild black hole in asymptotically
flat space is stable.\footnote{This stability is implied by classical
no-hair theorems, see for example \cite{Price}.  A more extensive list
of references on no-hair theorems can be found in \cite{ghpss}.  A
consequence of the present work is that these theorems cannot be
extended to charged black holes in AdS.}  Yet we conjecture that
\Conjecture\ gives a precise relation when the black object has a
non-compact translational symmetry.

Our focus in this paper is black holes in AdS and their black brane
limits; however the conjecture~\Conjecture\ is intended to apply
equally to any black brane.  It may apply even beyond the regime of
validity of classical gravity.  Any ``sensible'' gravitational
dynamics should satisfy the Second Law of Thermodynamics,
and~\Conjecture\ is motivated solely by intuition that Lorentzian time
evolution should proceed so as to increase the entropy.  (The
stipulation of translational invariance prevents finite volume effects
from vitiating simple thermodynamic arguments).  For instance, it has
recently been shown \cite{sahakian} that the near-extremal NS5-brane
has a negative specific heat arising from genus one contributions on
the string worldsheet (see also \cite{br,ho}, and \cite{kkk} for
related phenomena in 1+1-dimensional string theory).\footnote{We thank
D.~Kutasov for bringing \cite{sahakian,kkk} to our attention.}  This
is not classical gravity, but~\Conjecture\ leads us to expect an
instability in the Lorentzian time evolution of near-extremal
NS5-branes.\footnote{We thank M.~Rangamani for a number of discussions
on this point.}  The instability would drive the NS5-brane to a state
in which the energy density is non-uniformly distributed over the
world-volume.

\section{The $AdS_4$-RN solution and its cousins}
\label{Solutions}

The bosonic part of the lagrangian for ${\cal N}=8$ gauged
supergravity \cite{deWitOne,deWitTwo} in four dimensions involves the
graviton, 28 gauge bosons in the adjoint of $SO(8)$, and 70 real
scalars.  Because of the scalar potential introduced by the gauging
procedure, flat Minkowski space is not a vacuum solution of the
theory; rather, $AdS_4$ is.  It is known \cite{deWitThree} that the
maximally supersymmetric $AdS_4$ vacuum of ${\cal N}=8$ gauged
supergravity represents a consistent truncation of 11-dimensional
supergravity compactified on $S^7$.  The $AdS_4 \times S^7$ solution
can be obtained as the analytic completion of the near-horizon limit
of a large number of coincident M2-branes.\footnote{As stated in the
introduction, taking the number of M2-branes large makes the geometry
smooth on the Planck/string scale and thus suppresses corrections to
classical two-derivative gravity.}  Making the M2-branes near-extremal
corresponds to changing $AdS_4$ to the $AdS_4$-Schwarzschild solution.
Near-extremal M2-branes can also be given angular momentum in the
eight transverse dimensions.  There are four independent angular
momenta, corresponding to the $U(1)^4$ Cartan subgroup of $SO(8)$:
these reduce to electric charges in the $AdS_4$ description.  The
electrically charged black hole solutions can be obtained most
efficiently by first making a consistent truncation of the full ${\cal
N}=8$ gauged supergravity theory to the $U(1)^4$ gauge fields plus
three real scalars.  Consistent truncation means that any solution of
the reduced theory can be embedded in the full theory, with no
approximations.  For our purposes, it can be viewed as a sophisticated
technique for generating solutions.  The truncated bosonic lagrangian
is
  \eqn{NEightL}{\eqalign{
   &{\cal L} = {\sqrt{g} \over 2\kappa^2} \left[ R - 
    \sum_{i=1}^3 \left( {1 \over 2} (\partial\varphi_i)^2 +
     {2 \over L^2} \cosh\varphi_i \right) - 
     2 \sum_{A=1}^4 e^{\alpha^i_A \varphi_i} 
     (F_{\mu\nu}^{(A)})^2 \right]  \cr
   &\hbox{where} \qquad 
   \alpha^i_A = \pmatrix{ 1 & 1 & -1 & -1  \cr
                           1 & -1 & 1 & -1  \cr
                           1 & -1 & -1 & 1 } \,.
  }}
 We use the conventions of \cite{DuffLiu}, in particular, the metric
signature is $-$$+$$+$$+$ and $G_4 = 1$.  In \cite{DuffLiu} the
electrically charged solutions were found to be
  \eqn{DuffSoln}{\eqalign{
   ds^2 &= -{F \over \sqrt{H}} dt^2 + {\sqrt{H} \over F} dz^2 +
    \sqrt{H} z^2 d\Omega^2  \cr
   e^{2\varphi_1} &= {h_1 h_2 \over h_3 h_4} \quad
   e^{2\varphi_2} = {h_1 h_3 \over h_2 h_4} \quad
   e^{2\varphi_3} = {h_1 h_4 \over h_2 h_3}  \cr
   F_{0z}^{(A)} &= \pm {1 \over \sqrt{8} h_A^2} {Q_A \over z^2} \cr
   H &= \prod_{A=1}^4 h_A \quad 
   F = 1 - {\mu \over z} + {z^2 \over L^2} H \quad
   h_A = 1 + {q_A \over z}  \cr
   Q_A &= \mu \cosh\beta_A \sinh\beta_A \quad
   q_A = \mu \sinh^2 \beta_A
  }}
 where the signs on the gauge fields can be chosen independently.  We
will lose nothing by choosing them all to be $+$.  The quantities
$Q_A$ are the physical conserved charges, and they correspond to the
four independent angular momenta of M2-branes in eleven dimensions.
The mass is \cite{cgOne}
  \eqn{MassDef}{
   M = {\mu \over 2} + {1 \over 4} \sum_{A=1}^4 q_A \,,
  }
and the entropy is 
  \eqn{EntropyDef}{
   S = \pi z_H^2 \sqrt{H(z_H)}
  }
 where $z_H$ is the largest root of $F(z_H) = 0$.  Only for
sufficiently large $\mu$ do roots to this equation exist at all.  When
they don't, the solution is nakedly singular.

We will be most interested in the case where all four charges are
equal, $q_A = q$.  Then the solution can be written more conveniently
in terms of a new radial variable, $r=z+q$, and it takes the form
  \eqn{AdSRN}{\eqalign{
   ds^2 &= -f dt^2 + {dr^2 \over f} + r^2 d\Omega^2  \cr
   F_{0r} &= {Q \over \sqrt{8} r^2}  \cr
   f &= 1 - {2M \over r} + {Q^2 \over r^2} + {r^2 \over L^2} \,,
  }}
 with the scalars set to $0$.  In \AdSRN, $F_{0r}$ is the common value
of all four gauge field strengths $F_{0r}^{(A)}$.  The geometry
\AdSRN\ is a solution of pure Einstein-Maxwell theory with a
cosmological constant: it is the $AdS_4$-RN solution.

There are related solutions to maximally supersymmetric gauged
supergravity in five and seven dimensions, corresponding respectively
to spinning D3-branes and spinning M5-branes.  In the case of
D3-branes, there are six transverse dimensions, the rotation group is
$SO(6)$, the Cartan subalgebra is $U(1)^3$, and as a result there are
three independent angular momenta (or charges in the Kaluza-Klein
reduced description).  In the case of M5-branes, there are five
transverse dimensions, the rotation group is $SO(5)$, the Cartan
subalgebra is $U(1)^2$, and there are two independent angular
momenta/charges.  We will record here only the Einstein frame metric
in the Kaluza-Klein reduced description, in conventions where $G_N =
1$ and $L$ is the radius of the asymptotic AdS space.  For further
information on these solutions, the reader is referred to
\cite{cgOne,cejmOne}.  The metrics are
  \eqn{CousinMetrics}{\seqalign{\span\TC\quad & \span\TR}{
   AdS_5: & ds^2 = - H^{-{2 \over 3}} F dt^2 + H^{1 \over 3} 
    \left( {{dr^2} \over F} + r^2 d \Omega_3 \right) \cr
   &H = \prod_{A=1}^3 h_A \quad 
   F = 1 - {\mu \over r^2} + {r^2 \over L^2} H \quad
   h_A = 1 + {q_A \over r^2}  \cr
   AdS_7: & ds^2 = - H^{-{4 \over 5}}F dt^2 + H^{1 \over 5} 
    \left( {{dr^2} \over F} + r^2 d \Omega_3 \right) \cr
     &H = \prod_{A=1}^2 h_A \quad 
   F = 1 - {\mu \over r^4} + {r^2 \over L^2} H \quad
   h_A = 1 + {q_A \over r^4} \,. \cr
  }}

\section{Thermodynamics}
\label{Thermo}

\subsection{Generalities}
\label{GeneralThermo}

Given the solutions \DuffSoln\ and \CousinMetrics, we may read off the
entropy, the mass, and the conserved electric charges.  Typically it
is most straightforward to express these quantities in terms of the
non-extremality parameter $\mu$ and the boost parameters $\beta_A$.
However it is possible to eliminate $\mu$ and $\beta_A$ and find a
polynomial equation relating $M$, $S$, and the $Q_A$.  This equation
can be solved straightfowardly for $M$, but not in general for $S$.
We will quote explicit results for $M = M(S,Q_1,\ldots,Q_n)$ in the
next section.  In this section we will discuss thermodynamic stability
assuming that $M(S,Q_1,\ldots,Q_n)$ is known.

The microcanonical ensemble is usually specified by a function $S =
S(M,Q_1,\ldots,Q_n)$.  Assuming positive temperature (which is safe
for regular black holes since the Hawking temperature can never be
negative), one may always invert $M = M(S,Q_A)$ to $S = S(M,Q_A)$,
where now we abbreviate $Q_1,\ldots,Q_n$ to $Q_A$.  A standard claim
in classical thermodynamics is that the entropy for ``sensible''
matter must be concave down as a function of the other extensive
variables.  Locally this means that the Hessian matrix,
  \eqn{SHessian}{
   {\bf H}^S_{M,Q_A} \equiv 
    \pmatrix{ {\partial^2 S \over \partial M^2} &
               {\partial^2 S \over \partial M \partial Q_B}  \cr
              {\partial^2 S \over \partial Q_A \partial M} &
               {\partial^2 S \over \partial Q_A \partial Q_B} } \,,
  }
 satisfies ${\bf H}^S_{M,Q_A} \leq 0$, {\it i.e.} it has no positive
eigenvalues.  To understand what this requirement means, consider the
simplest case where $n = 0$ and $\partial^2 S / \partial M^2 > 0$.
This is the statement that the specific heat is negative.  A substance
with this property (in a non-gravitational setting, but equating mass
with energy) is unstable: if we start at temperature $T$, then it is
possible to raise the entropy without changing the total energy by
having some regions at temperature $T + \delta T$ and others at $T -
\delta T$.  Since we are implicitly assuming a thermodynamic limit, it
is irrelevant how big the domains of high and low temperature are.  In
a more refined description ({\it e.g.} Landau-Ginzburg theory), these
domains might have a preferred size, or at least a minimal size.

In the more general setting of many independent thermodynamic
variables, let us define intensive quantities
  \eqn{yDefs}{
   (y_0,y_1,\ldots,y_n) = (M/V,Q_1/V,\ldots,Q_n/V) \,,
  }
 where $V$ is the volume.  Suppose that ${\bf H}^S_{M,Q_A}$ has a
positive eigenvector: ${\bf H}^S_{M,Q_A} \vec{v} = \lambda \vec{v}$
with $\lambda > 0$.  Through a variation
  \eqn{yVars}{
   y_j \to y_j + \epsilon v_j \,,
  }
 where $\epsilon$ is a function of position which integrates to $0$,
we can raise the entropy without changing the total energy or the
conserved charges.  Thus positive eigenvectors of ${\bf H}^S_{M,Q_A}$
indicate the way in which mass density and charge density tend to
clump.  Presumably the eigenvector with the most positive eigenvalue
gives the dominant effect.

The stability requirement ${\bf H}^S_{M,Q_A} \leq 0$ may be rephrased
as ${\bf H}^M_{S,Q_A} \geq 0$, where ${\bf H}^M_{S,Q_A}$ is the
Hessian of $M$ with respect to $S$ and $Q_A$.  This is easy to
understand from a geometrical point of view.  ${\bf H}^S_{M,Q_A} \leq
0$ says that all the principle curvatures of $S(M,Q_A)$ point toward
negative $S$, or, equivalently, away from the point $(S,M,Q_A) =
(\infty,0,0)$.  Now, the point $(S,M,Q_A) = (0,\infty,0)$ is on the
opposite side of the co-dimension hypersurface defined by $S =
S(M,Q_A)$ from $(S,M,Q_A) = (0,\infty,0)$.  Thus all principle
curvatures should point toward $(0,\infty,0)$, which means that ${\bf
H}^M_{S,Q_A} \geq 0$.  To determine the region of thermodynamic
stability we may thus require $\det {\bf H}^M_{S,Q_A} > 0$, and then
take the smallest connected components around points which are known
to be stable.

While regions of stability are conveniently calculated from ${\bf
H}^M_{S,Q_A}$, it is not clear that the eigenvector of ${\bf
H}^S_{M,Q_A}$ with the largest positive eigenvalue can be read off
easily from ${\bf H}^M_{S,Q_A}$.  So it is useful to express ${\bf
H}^S_{M,Q_A}$ directly in terms of derivatives of $M(S,Q_A)$:
  \eqn{SMH}{\eqalign{
   {\partial^2 S \over \partial M^2} &= 
    -{1 \over (\partial M / \partial S)^3} 
      {\partial^2 M \over \partial S^2}  \cr
   {\partial^2 S \over \partial Q_A \partial M} &= 
     {1 \over (\partial M / \partial S)^3} \left[
      -{\partial M \over \partial S} 
        {\partial^2 M \over \partial Q_A \partial S} + 
       {\partial M \over \partial Q_A}
        {\partial^2 M \over \partial S^2} \right]  \cr
   {\partial^2 S \over \partial Q_A \partial Q_B} &= 
     {1 \over (\partial M / \partial S)^3} \Bigg[
      -\left( {\partial M \over \partial S} \right)^2 
      {\partial^2 M \over \partial Q_A \partial Q_B} - 
     {\partial^2 M \over \partial S^2}
      {\partial M \over \partial Q_A} {\partial M \over \partial Q_B} \cr
   &\qquad{} + 
     {\partial M \over \partial S} 
      \left( {\partial M \over \partial Q_A}
        {\partial^2 M \over \partial Q_B \partial S} + 
       {\partial M \over \partial Q_B}
        {\partial^2 M \over \partial Q_A \partial S}
      \right) \Bigg] \,.
  }}

\begin{figure}[h]
   \centerline{\psfig{figure=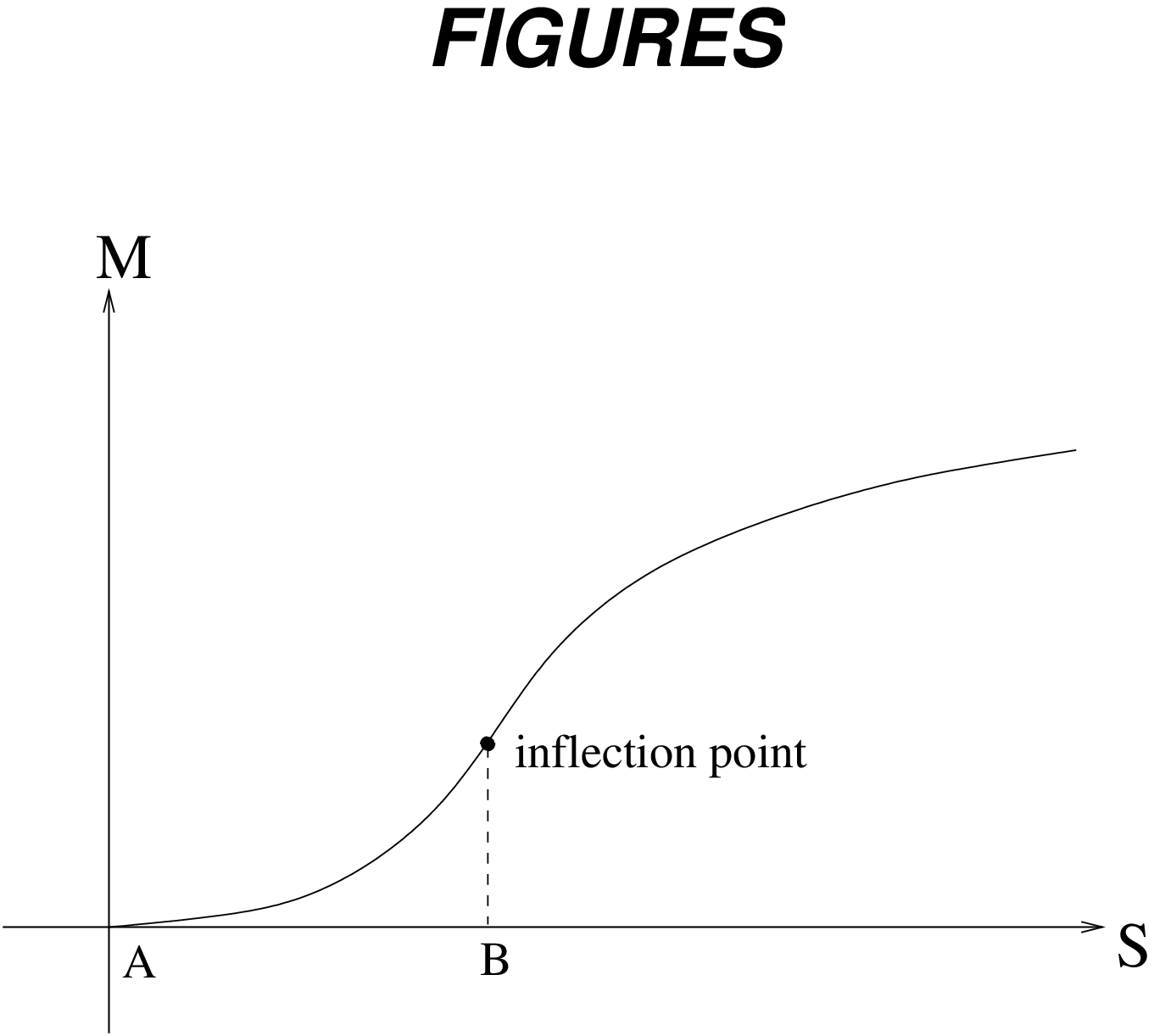,width=3in}}
   \caption{An example of a mass function whose convex hull is
flat.  The region we interpret as stable is from $A$ to $B$.}
   \label{figA}
  \end{figure}

A prescription for dealing with energy functions which violate the
convexity condition ${\bf H}^M_{S,Q_A} \leq 0$ is the Maxwell
construction, where one replaces $M(S,Q_A)$ with its convex hull (or
$S(M,Q_A)$ by its convex hull---it's the same thing).  This formal
procedure is equivalent to allowing mixed phases where some domains
have higher mass density or charge density than others.  The energy
functions resulting from charged black holes in AdS have the curious
property that the convex hull is completely flat in some directions,
so that chemical potentials (after taking the convex hull) are
everywhere zero.\fixit{Check} This arises because, in certain
directions, $M$ rises slower than any nontrivial linear function of
the other extensive variables.  In this situation the Maxwell
construction does not make much sense, because the mixed phases that
it calls for have charges and mass concentrated arbitrarily highly in
a small region, while the rest of the ``sample'' is at very low charge
and mass density.  A similar example in the simpler context of no
conserved charges would be a mass function $M(S)$ like the one in
figure~\ref{figA}.
 Here the natural physical interpretation is that the region between
$A$ and $B$ represents a stable phase, while the region to the right
of $B$ is unstable toward clumping most of its energy into small
regions.  This tendency would presumably be cut off by some minimal
length scale of domains.  The mass functions obtained from charged
black holes in AdS look roughly like figure~\ref{figA} along some
slices of the space of possible $(S,Q_A)$.  The interpretation we will
offer is that the black holes are stable in the regime of parameters
where convexity holds, and that they become dynamically unstable
toward clumping their charge and energy outside this
region.\footnote{There is a subtlety, discussed in \cite{cgTwo}, about
the precise location of the boundary between stable and unstable
regions.  As the system approaches the inflection point at $B$, finite
fluctuations might allow it to make small excursions into the unstable
region.  Working in a large $N$ limit where classical supergravity
applies on the AdS side of the duality seems to suppress such
fluctuations.}

The line of thought summarized in the previous paragraph was already
advanced in \cite{cgTwo}, but with only thermodynamic arguments to
support it.  A competing point of view was suggested in
\cite{Hawking}: the black holes in question have no ergosphere (more
precisely, there is Killing vector field which is timelike everywhere
outside the horizon), and this was argued to imply that there could be
no superradiant modes, and hence no classical instability in the
Lorentzian-time dynamics.  The argument used the dominant energy
condition, which need not always be satisfied by matter in AdS: in
fact, the scalars $\varphi_i$ in \NEightL\ violate the dominant energy
condition because of their tachyonic potential (which however does
satisfy the Breitenlohner-Freedman bound).

In \cite{gMitra}, an explicit numerical calculation demonstrated the
existence of a dynamical instability for certain $AdS_4$-RN black
holes (related to spinning M2-branes with all four spins equal, as
explained in the previous section).  We will discuss this calculation
at greater length in section~\ref{Instability}.  For now let us only
remark that in the limit of large black holes, where the horizon area
is infinite, the instability appears when thermodynamic stability is
lost, up to a discrepancy of $0.7\%$ which we suspect is numerical
error.  Furthermore, the combination of supergravity fields which
became unstable indicated a change in local charge densities precisely
in agreement with the analysis leading up to \yVars.  Thus the
conjecture \Conjecture\ was tested to reasonably good accuracy along a
two-parameter subspace (entropy and the common value of the four
charges) of the five-parameter phase space.  Further tests in $AdS_4$
are significantly more difficult because the metric usually enters in
to the perturbation equations in a non-trivial way.  However we will
indicate in section~\ref{Instability} another case where the metric
decouples.  Tests in $AdS_5$ and $AdS_7$ can also be performed most
easily in the equal charge case, but the analysis is somewhat more
tedious because the spinor formalism is not as well worked out in
higher dimensions (and probably is more cumbersome in any case).

Despite the absence of comprehensive tests, we will use \Conjecture\
and the idea that black hole perturbations should follow the most
unstable eigenvector of ${\bf H}^S_{M,Q_A}$ to propose in
section~\ref{Adiabatic} a qualitative picture of the evolution of
unstable black holes in AdS.  In brief, once the boundary of stability
is passed, the independent charges tend to clump separately, as if
they repelled one another but attracted themselves.  But this is only
an approximate tendency, with significant exceptions to be noted in
section~\ref{Adiabatic}. When a particular angular momentum density
becomes very large, it is possible that anti-de Sitter space fragments
by a classical process.\footnote{Fragmentation of AdS via tunneling
has been discussed in \cite{msBrill}.}  This also will be discussed at
greater length in section~\ref{Adiabatic}.  We emphasize that our
proposals for the evolution of unstable black holes are largely
conjectural, and difficult to check by any means other than numerical
solution of the full equations of motion.

\subsection{Explicit formulas}
\label{ExplicitThermo}

It is possible to eliminate all the auxiliary quantities from
\DuffSoln, \MassDef, and \EntropyDef, and express $M$ directly in
terms of the entropy and the physical charges as 
  \eqn{MExpress}{\eqalign{
   M ={1 \over {2\pi^{3 \over 2} L^2 \sqrt{S}}} \left[ {\prod_{A=1}^4({S^2 + \pi L^2 S + \pi^2 L^2 {Q_A}^2})}\right]^{1 \over 4}
  \,.
   }}
 We will often be interested in the limit of large black holes, $M/L
\gg 1$.  In this limit we have
  \eqn{FourCharge}{\eqalign{
    M = {1 \over {2\pi^{3 \over 2} L^2 \sqrt{S}}} \left[{\prod_{A=1}^4({S^2 + \pi^2 L^2 {Q_A}^2})}\right]^{1 \over 4}
  \,.
   }}
 with corrections suppressed by powers of $M/L$.  As $M/L$ approaches
infinity, one obtains a black brane solution in the Poincare patch of
$AdS_4$. Formally this limit can be taken by expanding \DuffSoln\ to
leading order in small $\beta_i$, dropping the 1 from $F$, and
replacing $S^2$ by $R^2$ in the metric.

\begin{figure}
   \centerline{\psfig{figure=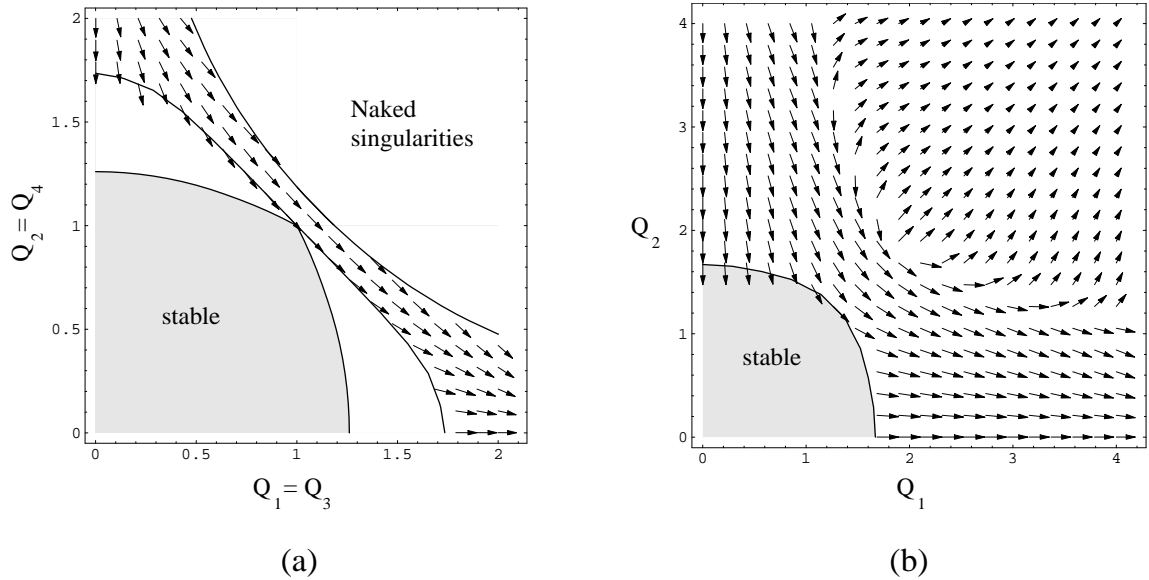,width=6in}}
   \caption{Plots of the most unstable eigenvector of the Hessian
matrix of $S(M,Q_1,Q_2,Q_3,Q_4)$.  The inner curves are boundaries of
stability.  The outer curves (when they are present) denote the
boundary between regular black branes and naked
singularities.}\label{figC}
  \end{figure}

As remarked in the previous section, local thermodynamic instability
can be expressed as convexity of the function $M(S,Q_1,Q_2,Q_3,Q_4)$.
By setting the Hessian of \FourCharge\ equal to zero, we obtain the
boundary separating the stable from the unstable region:
   \eqn{GenStabCurv}{\eqalign{
&   3S^8 - 2 \pi^2 L^2 S^6 \sum_{A=1}^4 Q_A^2 + \pi^4 L^4 S^4 \sum_{A < B} \left(Q_A Q_B \right)^2 - \pi^8 L^8 \prod_{A=1}^4 Q_A^2 = 0 \,.
   }}
First let us consider the case where the charges are pairwise set equal: $Q_1 = Q_3$ and $Q_2 = Q_4$. The above
equation then factorises, giving us three relevant factors:
   \eqn{TwoChargeInst}{\eqalign{
    \left(S^2 - \pi^2 L^2 Q_1^2 \right) \left(S^2 - \pi^2 L^2 Q_2^2 \right) \left(S^2 - {{\pi^2 L^2} \over 6} (Q_1^2 + Q_2^2 + \sqrt{Q_1^4 + Q_2^4 + 14 Q_1^2 Q_2^2})\right) = 0 \,.
}}
 When at least one of these factors become negative, ${\bf H}^M_{S,Q_A}$ develops a negative eigenvector and the black hole becomes thermodynamically unstable. A more convenient form may be obtained by eliminating $S$ in favor of
the mass $M$ and introducing the dimensionless variable $\chi_i = {Q_i
\over {M^{2 \over 3} L^{1 \over 3}}}$. The above three equations in the new variables become
   \eqn{LowerSym}{\eqalign{
    &\left[ \chi_1^4 + \chi_2^4 + 8 \chi_1^2 \chi_2^2 + (\chi_1^2 + \chi_2^2) \sqrt{\chi_1^4 + \chi_2^4 + 14 \chi_1^2 \chi_2^2} \right]^2  \cr
    &\qquad{} - 54 \left(\chi_1^2 + \chi_2^2 + \sqrt{\chi_1^4 + \chi_2^4 + 14 \chi_1^2 \chi_2^2}\right) = 0  \cr
    &\chi_1^6 + 2 \chi_1^2 \chi_2^2 (\chi_1^2 + \chi_2^2) - 4 = 0  
      \cr\noalign{\vskip1\jot}
    &\chi_2^6 + 2 \chi_1^2 \chi_2^2 (\chi_1^2 + \chi_2^2) - 4 = 0 \,.
}}
The region depicting thermodynamically stable black holes is the intersection of the areas under the 3 curves as shown in figure~\ref{figC}(a).
 
 The other relevant curve is the one separating nakedly singular
solutions from regular black holes.  The mathematical criterion for
having a regular black hole solution is that the polynomial $F$ in
\DuffSoln\ should have a zero.  In the large black hole limit, and in
terms of $\chi_1$ and $\chi_2$, this criterion reduces to
  \eqn{Horizon}{\eqalign{
  \chi_1^2 \chi_2^2 (\chi_1^8 + \chi_2^8) - 4 \chi_1^4 \chi_2^4 (\chi_1^4 + \chi_2^4) + 132 \chi_1^2 \chi_2^2 (\chi_1^2 + \chi_2^2) - 4 (\chi_1^6 + \chi_2^6)  + 6 \chi_1^6 \chi_2^6 - 432 =0 \,.
 }}
 To determine if a black hole with given values of mass and charges is
unstable, one first computes the values of $\chi_1$ and $\chi_2$ and
locates this point in figure~\ref{figC}(a). The black hole is unstable
if the point lies outside the shaded region depicting stable black
holes but is within the boundary which separates black holes with
naked singularities from those with a horizon. If the point lies in
the unshaded (unstable) region of the plot without the vector field
shown, it means that {\it within} each pair one charge wants to
increase while the other decreases. The unstable eigenvector has no
components along the hyperplane $Q_1=Q_3$ and $Q_2=Q_4$ and is not
shown.

Finally, let us collect the thermodynamic results for the special case
of all charges equal.  We see that thermodynamic instability is
present in the narrow region $1 < \chi < \sqrt{3}/2^{2/3}$.  The
associated eigenvector has the form $(0,1,-1,1,-1)$ where the
components are along the axes $M,Q_1,Q_2,Q_3,$ and $Q_4$ respectively:
it looks like one pair of charges wants to increase while the other
decreases.  This can happen only locally, with each of the four
charges conserved globally.

We'll also consider the case in which only two of the charges, $Q_1$
and $Q_2$, are non-zero. To get the region of thermodynamically stable
black holes, we set $Q_3 = Q_4 = 0$ in \GenStabCurv:
 \eqn{TwoZero}{\eqalign{
3S^4 - 2 \pi^2 L^2 S^2 (Q_1^2 + Q_2^2) + \pi^4 L^4 Q_1^2 Q_2^2 = 0
\,.
}}
 Just as we did in the previous case, we first eliminate $S$ in favor
of the mass $M$ and then introduce the dimensionless variables $\chi_i
= {Q_i \over {M^{2 \over 3} L^{1 \over 3}}}$ to get:
 \eqn{TwoZeroStab}{\eqalign{
  &10(\chi_1^6 + \chi_2^6) + 21 \chi_1^2 \chi_2^2 (\chi_1^2 + \chi_2^2)  \cr
  &\qquad{} + \left( 10\chi_1^4 + 10 \chi_2^4 + 26 \chi_1^2 \chi_2^2 \right) \sqrt{\chi_1^4 + \chi_2^4 - \chi_1^2 \chi_2^2} - 432 = 0 \,.
}}
 This is the boundary of the stable region, and is plotted in
figure~\ref{figC}(b).  Unlike the case of charges set equal pair-wise,
black holes with two charges set to zero always have a horizon.  This
may be connected with the fact that there is a limit of rotating
M2-branes with only two independent angular momenta nonzero which is a
well-defined multi-center M2-brane solution, while with all angular
momenta nonzero the corresponding limit is a singular configuration in
eleven dimensions \cite{klt,fgpwTwo,cglp}.

For black holes in $AdS_5$ and $AdS_7$, we will simply record here the
mass in terms of the entropy and charges:
  \eqn{FiveSeven}{\seqalign{\span\TC\quad & \span\TR}{
   AdS_5: & M = {3 \over 2 L^2 (2\pi^4 S)^{2/3}} 
    \left[ \prod_{A=1}^3 (4S^2 + \pi^4 L^2 Q_A^2) \right]^{1/3}  \cr
   AdS_7: & M = {5 \over 4 L^2 (4 \pi^9 S)^{2/5}}
    \left[ \prod_{A=1}^2 (16 S^2 + \pi^6 L^2 Q_A^2) \right]^{2/5} \,.
  }}
 Stability analyses similar to the $AdS_4$ case can be carried out for
$AdS_5$ and $AdS_7$.  Some work along these lines was presented in
\cite{cgTwo}, but the explicit expressions in \FiveSeven\ make the
calculations much easier.

\subsection{Adiabatic evolution}
\label{Adiabatic}

Tracking the evolution of unstable black holes in Lorentzian time is
difficult.  We have succeeded in establishing perturbatively the
existence of a dynamical instability for the very special case of all
charges equal: this is explained in section~\ref{Instability}.  This
simplest case required the numerical solution of a fourth order
ordinary differential equation with constraints at the horizon of the
black hole and the boundary of $AdS_4$.  Most other cases for black
holes in $AdS_4$ involve fluctuations of the metric, which makes the
analysis significantly harder.  To investigate the instabilities
beyond perturbation theory would require extensive numerical
investigation of the second order PDE's that comprise the equations of
motion of ${\cal N}=8$ gauged supergravity.

The aim of this section is to use thermodynamic arguments to guess the
qualitative features of the evolution of unstable black holes.  Here
we focus exclusively on the large black hole limit; however the
conclusions may remain valid to an extent for finite size black holes
with dynamical instabilities.  The intuition is that knowing the
entropy as a function of the other extensive parameters amounts to
knowing the zero-derivative terms in an effective Landau-Ginzburg
theory of the black hole (or of its dual field theory
representation).

As explained in the paragraph around \yDefs\ and \yVars, an unstable
eigenvector of ${\bf H}^S_{M,Q_A}$ (by which we mean one with positive
eigenvalue) suggests a direction in which a black hole solution can be
perturbed in order to raise entropy while keeping its total mass and
conserved charges fixed; moreover it was shown in \cite{gMitra} (as we
will explain in section~\ref{Instability}) that the black hole's
dynamical instability causes it to evolve in precisely the direction
that the eigenvector indicates.  The physics has no infrared cutoff,
as is typical in Gregory-Laflamme setups, so we may hope that the
charge and mass densities vary over long enough distance scales that
we may continue to use the most unstable eigenvector of ${\bf
H}^S_{M,Q_A}$ locally to determine the direction of the subsequent
evolution.  Following this line of thought to its logical conclusion
leads us to the claim that the mass density and charge densities will
locally evolve, subject to the constraints of conserving total energy
and charge, from their initial values to values along a characteristic
curve of the unstable vector field of ${\bf H}^S_{M,Q_A}$.  This can
only be approximately correct: finite wavelength distortions will
occur, and it is not precisely right anyway to say that the
time-evolution of Einstein's equations proceeds so as to maximize
black hole entropy.  Nevertheless it seems to us likely that a correct
qualitative picture will emerge from tracking the flows generated by
the most unstable eigenvector of ${\bf H}^S_{M,Q_A}$.  At late times,
or when charge and mass density are highly concentrated in small
regions, another description is needed.

The characteristic curves of the most unstable eigenvector of ${\bf
H}^S_{Q,M_A}$ may terminate in a region of stability, or in a region
of naked singularities.  Cosmic Censorship plus the conjectures of the
previous paragraph suggest that the latter should never happen.  This
can be checked explicitly for the examples that we have.  To this end,
one can choose a generic value of charges and mass so that the black
hole is almost naked, then determine the most unstable eigenvector of
${\bf H}^S_{M,Q_A}$, and then check that it is tangent to the surface
separating naked singularities from regular black holes.  We carried
this out numerically for several cases and verified tangency; however
we do not have a general argument.  It appears, in fact, that the
normal vector to the surface separating naked singularities from black
holes is a stable eigenvector of ${\bf H}^S_{M,Q_A}$ ({\it i.e.} its
eigenvalue is negative)---at least in the three-dimensional subspace
with $Q_1 = Q_3$ and $Q_2 = Q_4$---so the obvious approach to an
analytic demonstration that Cosmic Censorship is not violated by
adiabatic evolution of black holes is to show that this normal vector
is always a stable eigenvector of ${\bf H}^S_{M,Q_A}$.  For now we
content ourselves with the observation that in all the cases we have
checked numerically, adiabatic evolution does stay in the region of
regular black holes.

It is also possible that a characteristic curve becomes unstable at
some point, in the sense that nearby characteristic curves diverge
from it.  To refine our previous claim, we may suppose that the black
hole evolves along a bundle of nearby characteristic curves emanating
from the original mass and charge density.  This bundle may remain
nearly one-dimensional, or it may split or become higher dimensional.
We will not investigate the stability properties of the characteristic
curves in any detail.  Note that we are not attempting to specify any
spatial or temporal properties of the evolution, only the range of
mass and charge densities which form.

We present in figure~\ref{figC} plots of unstable eigenvectors of the
Hessian matrix ${\bf H}^S_{M,Q_A}$, projected onto a plane
parametrized by two of the charges.
 From these vector fields, we may conclude that the different charges
exhibit some tendency to separate from one another, but that this does
not always happen, as in the upper right part of figure~\ref{figC}(b).
The crucial point is that the unstable eigenvectors don't have a
component normal to the boundary between naked singularities and
regular black holes. Although this appears obvious from
figure~\ref{figC}(a), the plot is slightly misleading in that the
eigenvectors have been projected onto the plane of $Q_1=Q_3$ and
$Q_2=Q_4$.  One must preserve the components of vectors in the $M$
direction to verify tangency.

When some angular momenta become large compared to the entropy for a
spinning M2-brane solution, the geometry in eleven dimensions is
approximately given by a rotating multi-center brane solution
\cite{klt}.  If one angular momentum is large, this multi-center
solution is in the shape of a disk; if two are large and equal, it has
the shape of a filled three-sphere.  It seems clear that solutions of
this form in an asymptotically flat eleven-dimensional spacetime are
unstable toward fragmentation in the directions transverse to the
M2-brane.  This would mean that anti-de Sitter space fragments.  In
terms of the $SU(N)$ gauge theory, the disk corresponds to a
$U(1)^{N-1}$ Higgsing, and in the fragmentation process some groups of
$U(1)$'s try to come together to partially restore gauge invariance.
It is not certain that such fragmentation occurs, particularly if the
angular momentum density is large only locally.  We merely indicate it
as a possibility in the complicated late-time evolution of unstable
black holes.

Finally, it is possible that the unstable black holes evolve to a some
new stationary solution, presumably with a non-uniform horizon.  No
such solutions are known.  It seems to us most likely that, upon
becoming unstable, black holes in AdS undergo an evolution which
eventually produces large curvatures.

\section{Existence of a dynamical instability}
\label{Instability}

The existence of dynamical instabilities for thermodynamically
unstable black branes should be completely generic.  However, as
mentioned already, the stability analysis is technically complicated
for the general case of unequal charges: perturbations of the metric,
four gauge fields, and three scalars lead to difficult coupled partial
differential equations.  Here we focus on the $AdS_4$-RN example,
where the metric decouples and the problem can be reduced to a single
gauge field and a single scalar.  A formal argument relating
thermodynamic and dynamical instability was suggested in
\cite{gMitra}, using the identification of the free energy with the
Euclidean supergravity action; however we have not yet succeeded in
making this argument rigorous.

Because the unstable eigenvectors of ${\bf H}^S_{M,Q_A}$ (for all
charges equal and sufficiently large) do not involve any change in the
mass density, it is natural to expect that the perturbations that give
rise to an unstable mode do not involve the metric.\footnote{Indeed,
we suspect that the decoupling of the metric is possible precisely
when there is an eigenvalue of ${\bf H}^S_{M,Q_A}$ which does not have
a component in the $M$ direction.}  More precisely, because of the
form of the unstable eigenvectors, we expect that a relevant
perturbation is
  \eqn{deltaF}{
   \delta F_A = \alpha^i_A \delta F
  }
 for some $\delta F$ and fixed $i$, where the $\alpha^i_A$ were
defined in \NEightL.  In section~\ref{Thermo} we saw explicitly that
$\delta Q_1 = \delta Q_3 = -\delta Q_2 = -\delta Q_4$ gave an unstable
eigenvector; now we make a trivial alteration and focus on $\delta Q_1
= \delta Q_2 = -\delta Q_3 = -\delta Q_4$.  Correspondingly we set
$i=1$ in \deltaF.

The spectrum of linear perturbations to charged black holes in AdS has
been considered before \cite{cort}, but for the most part the
perturbations under study were minimally coupled scalars.  It is
impractical to sift through the entire spectrum of supergravity
looking for unstable modes (or tachyonic glueballs, in the language of
\cite{cort}).  The point of the previous paragraphs is that
thermodynamics provides guidance not only on when to expect an
instability, but also in which mode.

It is straightforward to start with the lagrangian in \NEightL\ and
show that linearized perturbations to the equations of motion result
in the following coupled equations:
  \eqn{CoupledScalar}{\eqalign{
   & d\delta F = 0 \qquad 
     d*\delta F + d\delta\varphi_1 \wedge *F = 0  \cr
   & \left[ \square + {2 \over L^2} - 8 F_{\mu\nu}^2 \right] 
      \delta\varphi_1 - 16 F^{\mu\nu} \delta F_{\mu\nu} = 0 \,.
  }}
 Here $\square = g^{\mu\nu} \nabla_\mu \partial_\nu$ is the usual
scalar laplacian.  $F$ in \CoupledScalar\ is the background field
strength in \AdSRN: it is the common value of the four $F_A$.  $\delta
F$ is {\it not} the variation in $F$ itself; rather, the variation of
the $F_A$ is expressed in terms of $\delta F$ in \deltaF, with $i=1$.
The variation of the field strength is in a direction {\it orthogonal}
to the background field strength of the $AdS_4$-RN solution.  The
graviton decouples from the linearized perturbation equations: $\delta
T_{\mu\nu}$ vanishes at linear order in $\delta F$ because $\delta F_A
\cdot F_A = 0$.\footnote{Besides the all-charges-equal case, we know
of one other case where the metric decouples at linear order: $Q_1 =
Q_3$ with $Q_2 = Q_4 = 0$.  There may be other cases as
well---presumably whenever $Q_A \cdot \delta Q_A = 0$ and $\delta S =
0$ for an unstable eigenvector $(\delta S,\delta Q_A)$ of ${\bf
H}^M_{S,Q_A}$.}

For comparison, we write down the linearized equations for
fluctuations of the other scalars:
  \eqn{UncoupledScalars}{
   \left[ \square + {2 \over L^2} - 8 F_{\mu\nu}^2 \right]
    \delta\varphi_i = 0
  }
 for $i=2,3$.  It was shown in \cite{Hawking} that any perturbation
involving only matter fields satisfying the dominant energy condition
could not result in a normalizable unstable mode (that is, a
normalizable mode which grows exponentially in Lorentzian time).  It was
conjectured \cite{Hawking,HawkingStrings} that in fact there was no
classical instability at all.  The scalars $\varphi_i$ do not satisfy
the dominant energy condition because of the potential term in
\NEightL.  Thus the outcome of our calculations is not fore-ordained by
general arguments, and we have a truly non-trivial check on the
classical stability of highly charged black holes in ${\cal N}=8$
gauged supergravity.  In fact, our results turn out to be in conflict
with the claim of classical stability in
\cite{Hawking,HawkingStrings}.

Decoupling the equations in \CoupledScalar\ is a chore greatly
facilitated by the use of the dyadic index formalism introduced in
\cite{NP}.  For the reader interested in the details, we present an
outline of the derivation in section~\ref{Dyadic}.  The final result
is the fourth order ordinary differential equation (ODE)
  \eqn{FinalODE}{\eqalign{
   &\left( {\omega^2 \over f} + \partial_r f \partial_r - 
    {\ell (\ell+1) \over r^2} \right) r^3 
    \left( {\omega^2 \over f} + \partial_r f \partial_r - 
    {\ell (\ell+1) \over r^2} - {2M \over r^3} + 
    {4 Q^2 \over r^4} \right) r \delta\tilde\varphi_1(r) =  \cr
    &\qquad\quad 4 Q^2 \left( {\omega^2 \over f} + 
     \partial_r f \partial_r \right) \delta\tilde\varphi_1(r) \,,
  }}
 where we have assumed the separated form $\delta\varphi_1 = \Re
e^{-i\omega t} Y_{\ell m} \delta\tilde\varphi_1(r)$, where $Y_{\ell
m}$ is the usual spherical harmonic on $S^2$.  This is to be compared
with the separated equation for the other scalars:
  \eqn{SecondODE}{\eqalign{
   \left( {\omega^2 \over f} + \partial_r f \partial_r - 
    {\ell (\ell+1) \over r^2} - {2M \over r^3} + 
    {4 Q^2 \over r^4} \right) r \delta\tilde\varphi_i(r) = 0
  }}
 for $i=2,3$.

\subsection{Dyadic index derivation of \FinalODE}
\label{Dyadic}

To derive \FinalODE\ using the dyadic index formalism, it is
convenient first to switch to $+$$-$$-$$-$ signature to avoid sign
incompatibilities between the raising and lowering of dyadic and
vector indices.  One introduces a null tetrad of vectors,
$(l^\mu,n^\mu,m^\mu,\bar{m}^\mu)$, defined so that $l^\mu n_\mu =
-m^\mu m_\mu = 1$ and all other inner products vanish.  Next define
  \eqn{SigmaDef}{
   \sigma^\mu_{\Delta\dot\Delta} = 
    \pmatrix{ l^\mu & m^\mu  \cr  \bar{m}^\mu & n^\mu }
  }
 and set $D = l^\mu \partial_\mu$, $\Delta = n^\mu \partial_\mu$,
$\delta = m^\mu \partial_\mu$, $\bar\delta = \bar{m}^\mu
\partial_\mu$.  Vector indices are converted into dyadic indices by
setting $v_{\Delta\dot\Delta} = \sigma^\mu_{\Delta\dot\Delta} v_\mu$.
Dyadic indices are raised and lowered using northwest contraction
rules with $\epsilon_{01} = \epsilon^{01} = \epsilon_{\dot{0}\dot{1}}
= \epsilon^{\dot{0}\dot{1}} = 1$.  By demanding that
$\sigma^\mu_{\Delta\dot\Delta}$ is covariantly constant, one can
obtain a unique covariant derivative $D_\mu$, whose action on a spinor
is
  \eqn{DmuDef}{
   D_\mu \psi_\Gamma = \partial_\mu \psi_\Gamma - \psi_\Sigma
    \gamma_\mu{}^\Sigma{}_\Gamma \,.
  }
 The so-called spin coefficients,
$\gamma_{\Delta\dot\Delta\Sigma\Gamma} = \sigma^\mu_{\Delta\dot\Delta}
\gamma_{\mu\,\Sigma\Gamma}$, are conventionally written as
  \eqn{DyadNotation}{\seqalign{\strut\span\TL & \span\TR \qquad & 
     \span\TL & \span\TR}{
   \gamma_{0\dot{0}\Sigma\Gamma} &= 
    \pmatrix{ \kappa & \epsilon \cr \epsilon & \pi } &
   \gamma_{0\dot{1}\Sigma\Gamma} &=
    \pmatrix{ \sigma & \beta \cr \beta & \mu } \cr
   \gamma_{1\dot{0}\Sigma\Gamma} &=
    \pmatrix{ \rho & \alpha \cr \alpha & \lambda } &
   \gamma_{1\dot{1}\Sigma\Gamma} &=
    \pmatrix{ \tau & \gamma \cr \gamma & \nu } \,.
  }}
 A less compressed presentation of dyadic index formalism can be found
in \cite{NP,Wald}, and the appendix to \cite{gdyad}.  

For $AdS_4$-RN, a convenient choice of the null tetrad and the
corresponding nonzero spin coefficients are as follows:
  \eqn{Kinnersley}{\seqalign{\span\TL & \span\TR \qquad & \span\TL
    & \span\TR}{
   l^\mu &= (1/f,1,0,0) & 
    n^\mu &= {1 \over 2} (1,-f,0,0)  \cr
   m^\mu &= {1 \over r \sqrt{2}} (0,0,1,i \csc\theta) &
    \bar{m}^\mu &= {1 \over r \sqrt{2}} (0,0,1,-i \csc\theta)
  }}
  \eqn{SpinCoefs}{
   \rho = -{1 \over r} \quad \mu = -{f \over 2r} \quad
   \gamma = {f' \over 4} \quad 
   \alpha = -\beta = -{\cot\theta \over \sqrt{8} r} \,.
  }
 in \Kinnersley\ and \SpinCoefs\ we have not yet taken the black brane
limit.  Taking this limit replaces $\csc\theta$ by $1$ in \Kinnersley\
and sets $\alpha=\beta=0$ in \SpinCoefs.  Proceeding without the black
brane limit, we trade the real antisymmetric tensor $F_{\mu\nu}$ for a
complex symmetric tensor,
  \eqn{PhiDef}{
   \Phi^{(0)}_{\Delta\Gamma} = 
    \pmatrix{ \phi_0^{(0)} & \phi_1^{(0)}  \cr
              \phi_1^{(0)} & \phi_2^{(0)} }
  }
 through the formula
  \eqn{FPhi}{
   4\sqrt{2}
   F_{\mu\nu} \sigma^\mu_{\Delta\dot\Delta} \sigma^\nu_{\Gamma\dot\Gamma} = 
    \Phi^{(0)}_{\Delta\Gamma} \epsilon_{\dot\Delta\dot\Gamma} +
    \bar\Phi^{(0)}_{\dot\Delta\dot\Gamma} \epsilon_{\Delta\Gamma} \,.
  }
 The factor of $4\sqrt{2}$ in \PhiDef\ is for convenience: the
$AdS_4$-RN background has $\phi_1^{(0)} = Q/r^2$ and all other
components zero.  In the same way we trade in $\delta F_{\mu\nu}$ for
$\Phi_{\Delta\Gamma}$, whose components are $\phi_0$, $\phi_1$, and
$\phi_2$, with a similar factor of $4\sqrt{2}$.  Finally, we write
$\varphi$ in place of $\delta\varphi_1$ to avoid the ambiguity in the
meaning of $\delta$.  

The first order equations for the gauge field in \CoupledScalar\ can
now be cast in dyadic form as follows:
  \eqn{DyadicF}{
   D^\Delta{}_{\dot\Gamma} \Phi_{\Delta\Gamma} + 
    {1 \over 2} \partial^{\Delta\dot\Delta} \varphi
    (\Phi^{(0)}_{\Delta\Gamma} \epsilon_{\dot\Delta\dot\Gamma} + 
     \bar\Phi^{(0)}_{\dot\Delta\dot\Gamma} \epsilon_{\Delta\Gamma}) = 0 \,.
  }
 In components, these equations read
  \eqn{DFComp}{\eqalign{
   (D-2\rho) \phi_1 - (\bar\delta-2\alpha) \phi_0 &= 
     -\phi_1^{(0)} D\varphi  \cr
   (\Delta+\mu-2\gamma) \phi_0 - \delta\phi_1 &= 0  \cr
   (D-\rho) \phi_2 - \bar\delta\phi_1 &= 0  \cr
   (\delta+2\beta) \phi_2 - (\Delta+2\mu) \phi_1 &= 
     \phi_1^{(0)} \Delta\varphi \,.
  }}
 It is possible to combine these equations into three second order
equations in which only a single $\phi_i$ appears.  Together with the
scalar equation, these equations are equivalent to \CoupledScalar:
  \eqn{SecondPass}{\eqalign{
   \left[ (D-3\rho)(\Delta+\mu-2\gamma) - 
    \delta(\bar\delta-2\alpha) \right] \phi_0 &= 
    -\phi_1^{(0)} \delta D\varphi  \cr
   \left[ (\Delta+3\mu)(D-\rho) - 
    \bar\delta(\delta+2\beta) \right] \phi_2 &=
    -\phi_1^{(0)} \bar\delta\Delta\varphi  \cr
   \left[ (D-2\rho)(\Delta+2\mu) - 
    (\delta+\beta-\alpha)\bar\delta \right] \phi_1 &= 
    -\phi_1^{(0)} D\Delta\varphi  \cr
   \left[ \square + {2 \over L^2} + 2 (\phi_1^{(0)})^2 \right] \varphi &=
    -4 \phi_1^{(0)} \Re\phi_1
  }}
 where we have made use of the fact that the spin coefficients are all
real for $AdS_4$-RN.  The equations \DFComp\ are a special case of
(3.1)-(3.4) of \cite{Teukolsky}.  The first and second equations of
\SecondPass\ are (3.5) and (3.7) of \cite{Teukolsky}, and the third is
derived in a similar manner.  The fourth is the scalar equation in
\CoupledScalar, but to preserve the definition of $\square$ we write
$\square = -g^{\mu\nu} \nabla_\mu \partial_\nu$ in $+$$-$$-$$-$
conventions.  The differential operators in the third equation of
\SecondPass\ are purely real (this takes a bit of checking for
$(\delta+\beta-\alpha)\bar\delta$), so we can take the real and
imaginary parts of this equation.  The equation for $\Im\phi_1$
decouples from all the others.  The equations for $\phi_0$ and
$\phi_2$ are sourced by $\varphi$, but $\phi_0$ and $\phi_2$ do not
otherwise enter; thus one can solve first for $\Re\phi_1$ and
$\varphi$, and afterwards use the first and second equations in
\SecondPass\ to obtain $\phi_0$ and $\phi_2$.  Since $\phi_1^{(0)}$ is
nowhere vanishing, the last equation in \SecondPass\ can be used to
eliminate $\Re\phi_1$ algebraically.  The final result is
  \eqn{GFODE}{
   \left[ (D-2\rho)(\Delta+2\mu) - (\delta+\beta-\alpha)\bar\delta \right]
    {1 \over 4\phi_1^{(0)}} \left[ \square + {2 \over L^2} + 
    2(\phi_1^{(0)})^2 \right] \varphi = \phi_1^{(0)} D\Delta\varphi \,.
  }
 Plugging in the separated ansatz $\varphi = \Re\left\{ e^{-i\omega t}
Y_{\ell m} \delta\tilde\varphi_1(r) \right\}$, one easily obtains
\FinalODE.

\subsection{Numerical results from the fourth order equation}
\label{Numerics}

A dynamical instability exists if there is a normalizable, unstable
solution to \FinalODE\ or to \SecondODE.  Neither of these equations
admits a solution in closed form, so we have resorted to numerics.
Briefly, the conclusion is that, in the black brane limit and within
the limits of numerical accuracy, we find a single unstable mode for
\FinalODE\ precisely when $\chi > 1$, and no instabilities for
\SecondODE.  This is completely in accord with the intuition from
thermodynamics: \SecondODE\ represents a fluctuation that has nothing
to do with the variation of charges that gave the unstable eigenvector
of the Hessian matrix of $M(S,Q_1,Q_2,Q_3,Q_4)$.  The unstable mode in
\FinalODE\ persists to finite size $AdS_4$-RN black holes, but
eventually disappears for small enough black holes.

To carry out a numerical study of \FinalODE, the first step is to cast
the equation in terms of a dimensionless radial variable $u$, a
dimensionless charge parameter $\chi$, a dimensionless mass parameter
$\sigma$, and a dimensionless frequency $\tilde\omega$:
  \eqn{uEtaDef}{
   u = {r \over M^{1/3} L^{2/3}} \qquad 
   \chi = {Q \over M^{2/3} L^{1/3}} \qquad
   \sigma = \left( {L \over M} \right)^{2/3} \qquad
   \tilde\omega = {\omega L^{4/3} \over M^{1/3}} \,.
  }
 Then we have 
  \eqn{FourthForm}{\eqalign{
   & \left( {\tilde\omega^2 \over \tilde{f}} + 
     \partial_u \tilde{f} \partial_u - 
     \sigma {\ell (\ell+1) \over u^2} \right)
     u^3 
    \left( {\tilde\omega^2 \over \tilde{f}} + 
     \partial_u \tilde{f} \partial_u - 
     \sigma {\ell (\ell+1) \over u^2} - 
      {2 \over u^3} + {4\chi^2 \over u^4} \right)
     u \delta\tilde\varphi_1 =  \cr
   & \qquad\quad 4 \chi^2 \left( {\tilde\omega^2 \over \tilde{f}} + 
     \partial_u \tilde{f} \partial_u \right) \delta\tilde\varphi_1  \cr
   & f = \sigma - {2 \over u} + {\chi^2 \over u^2} + u^2 \,.
  }}
 Evidently, the dimensionless control parameters are $\ell$ (the
partial wave number), $\sigma$, and $\chi$.  Using Mathematica, we
solved \FourthForm\ numerically via a shooting method, and obtained
wavefunctions $\delta\tilde\varphi_1(r)$ which fall off like $1/r^2$
near the boundary of $AdS_4$ and at least as fast as
$(r-r_H)^{|\omega|/f'(r_H)}$ near the horizon.

\begin{figure}
   \centerline{\psfig{figure=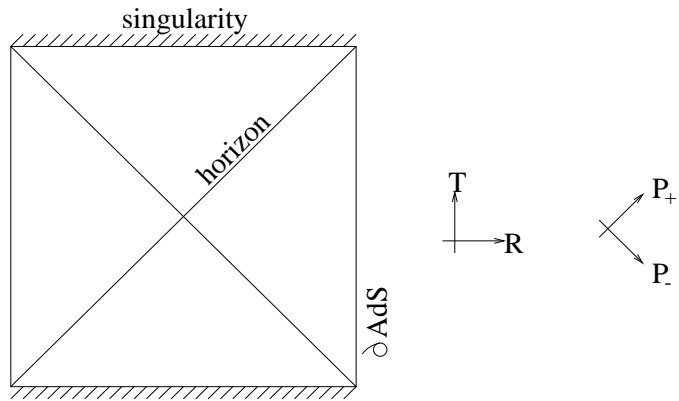,width=3.5in}}
   \caption{The Penrose diagram of a regular AdS black hole.  We can
take $T=R=P_+=P_-=0$ at the center of the diagram.  The black hole
horizon is the diagonal line going up and right from the
origin.}\label{figB}
  \end{figure}

To check that the wavefunction is well behaved near the
horizon\footnote{We thank G.~Horowitz for suggesting that this check
should be made.} let us transform to Kruskal coordinates.  The metric
near the horizon is
  \eqn{Kruskal}{\eqalign{
   ds^2 \approx -f'(r_H) (r-r_H) dt^2 + {dr^2 \over f'(r_H)(r-r_H)} + 
     r_H^2 d\Omega_2^2 \,,
  }}
 where $r_H$ is the radius of the horizon.  Dropping the $S^2$ piece
and introducing a tortoise coordinate $r_*$, null coordinates
$P_{\pm}$, and Kruskal coordinates $(T,R)$ according to 
  \eqn{NewCoords}{\eqalign{
   {dr_* \over dr} &= {1 \over f'(r_H) (r-r_H)}  \cr
   P_{\pm} &= e^{{1 \over 2} f'(r_H) (\pm t + r_*)} 
    = \pm T + R \,,
  }}
 one finds that the near-horizon metric is indeed regular:
  \eqn{NearHMetric}{
   ds_2^2 = -f'(r_H) (r-r_H) dt^2 + {dr^2 \over f'(r_H)(r-r_H)}
    = {4 \over f'(r_H)} (-dT^2 + dR^2) \,.
  }
 Having a radial wavefunction $\delta\tilde\varphi_1(r) =
(r-r_H)^{|\omega|/f'(r_H)} \rho(r-r_H)$, where $\rho(r-r_H)$ remains
bounded at the horizon, means that the time-dependent perturbation
(with angular dependence suppressed) is 
  \eqn{VarphiAsymptotics}{
   \delta\varphi_1(t,r) \sim (r-r_H)^{|\omega|/f'(r_H)}
    e^{|\omega| t} \rho(r-r_H) \sim 
    P_+^{2|\omega|/f'(r_H)} \rho(P_+ P_-) \,,
  }
 which remains bounded as $P_- \to 0$.  The black hole horizon is at
$P_- = 0$, $P_+ > 0$ (see figure~\ref{figB}).  Thus we see that the
perturbation is small at the horizon in good coordinates, at least for
small $P_+$.  (As the perturbation grows, the horizon eventually
starts to fluctuate, but this is not an issue in the question of
whether the instability exists).

\begin{figure}
   \centerline{\psfig{figure=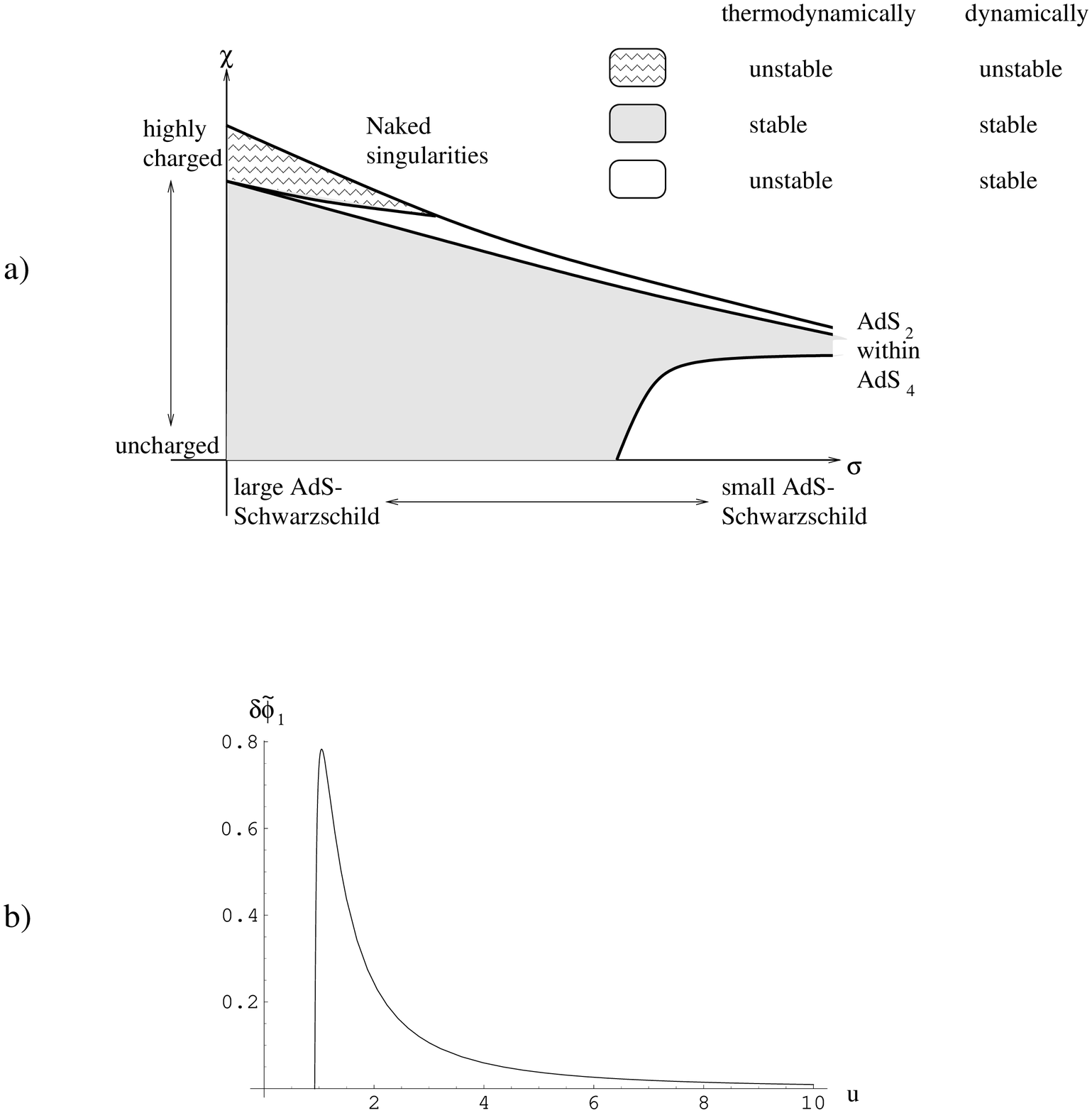,width=4.9in}}
   \caption{(a) A topologically correct representation of dynamical
and thermodynamic stability in the whole $\chi$-$\sigma$ plane (but
see the text regarding possible overlap of the two shaded regions).
(b) A sample normalizable wave-function with negative $\omega^2$: here
$\sigma = 0.3$, $\chi = 0.96$, and $\tilde\omega^2 =
-0.281$.}\label{figD}
  \end{figure}

A qualitative summary of our numerical results is displayed in
figure~\ref{figD}(a).  An example of a normalizable wave-function with
negative $\omega^2$ is shown in figure~\ref{figD}(b).  Some points to
note are:
  \begin{itemize}
  \item The boundary of the region of dynamical stability comes from
instability in the $\ell = 1$ mode.  The $\ell = 0$ mode is projected
out by charge conservation.  Higher $\ell$ modes become unstable in
the upper left part of the shaded triangle in figure~\ref{figD}(b).
The boundaries of dynamical instability for different $\ell$ all come
together at $\sigma = 0$.

  \item At $\sigma = 0$, thermodynamic stability is lost at $\chi =
1$, whereas dynamical instability sets in at $\chi = 1.007$.  We
believe that the $0.7\%$ discrepancy is due to numerical error.

  \item We have drawn the regions of dynamical instability and
thermodynamic stability as disjoint in figure~\ref{figD}(a).  In fact,
our current numerics shows them overlapping by about $0.1\%$ around
$\sigma = 0.1$.  We do not view this as significant because the
numerical errors seem to be around $1\%$.
  \end{itemize}

Finally, it is worth pointing out that the string theory program of
computing black hole entropy via a microscopic state count in a field
theory dual (see for example \cite{sv}, or \cite{PeetTASI} for a
review) has proved hard to extend past the boundaries of thermodynamic
stability.  For instance, we have a good understanding of the entropy
of near-extremal D3-branes \cite{gkPeet,AndyUnp}, but not of small
Schwarzschild black holes in AdS.  It seems to us that this is no
accident: most sensible field theories have log-convex partition
functions, and this translates into Hessian matrices ${\bf
H}^S_{M,Q_A}$ which have no negative eigenvalues.  Pushing past the
boundary of thermodynamic stability in a field theory may be possible
(particularly as one crosses a phase boundary and begins to nucleate
the new phase), but doing so seems likely to produce dynamical
instabilities in the Lorentzian time-evolution.  This point of view
has indeed informed our entire investigation.  

A dual field theory description of a small Schwarzschild black hole in
AdS must involve thermodynamic instability but no dynamical
instabilities.  We believe that finite volume effects in the field
theory are essential in this regard: if one imagines a Landau-Ginzburg
effective description of the field theory, then derivative terms must
restore stability to a system whose infrared tendencies are controlled
by the thermodynamic instability.  Various properties of small
AdS-Schwarzschild black holes have been explored (see for example
\cite{hhOne,hhTwo}), but the basic problem of reconciling
thermodynamic instability with dynamical stability in the presence of
a field theory dual remains to be addressed.

\section{Conclusions}
\label{Conclude}

A common conception of the Gregory-Laflamme instability is that a
uniform solution to Einstein's equations (plus matter) competes with a
non-uniform solution, and the non-uniform solution sometimes wins out
entropically.  In such a situation, the generic expectation is that
there is a first order tunneling transition from the uniform to the
non-uniform state, which may take place very slowly due to a large
energetic barrier.  In fact, the original papers \cite{glOne,glTwo}
focused mainly on demonstrating the existence of unstable modes in a
linearized perturbation analysis of the uniform solution.  The
distinction is between global and local stability.  At the level of
classical gravity/field theory, the latter concept is more meaningful,
because with quantum effects suppressed it is impossible to tunnel
away from a locally stable solution.  The aim of this paper and its
shorter companion paper \cite{gMitra} has been to study local
dynamical stability of black holes in anti-de Sitter space in relation
to a particular notion of local thermodynamic stability, namely
downward concavity of the entropy as a function of the other extensive
variables.  We reach two main conclusions:
  \begin{enumerate}
   \item In the limit of large black holes in AdS, dynamical and
thermodynamic stability coincide.  This conclusion is supported by
numerical evidence.  The small discrepancy between the observed onset
of dynamical and thermodynamic instabilities is probably numerical
error.

  \item Dynamical instabilities persist for finite size black holes in
AdS, down to horizon radii on the order of the AdS radius.  The
evidence is again only numerical, but we believe the final answer is
correct and robust.
  \end{enumerate}

We regard point~1 as a partial verification of a rather more general
conjecture, namely that black branes should have Gregory-Laflamme
instabilities (in the local, dynamical sense of the original papers
\cite{glOne,glTwo}) precisely when thermodynamic stability is lost.

Point~2 is surprising because it is the first known example of a
stationary black hole solution with a point-like singularity which
exhibits a dynamical Gregory-Laflamme instability.  Furthermore, it
shows that no-hair theorems cannot always hold in anti-de Sitter
space.  Black branes which experience a Gregory-Laflamme instability
are often supposed to split their horizons and fall into pieces.  For
this to happen to a horizon whose topology is ${\bf S}^2$ and which
cloaks a point-like singularity would be truly novel: what could tear
apart a point-like singularity?

Is Cosmic Censorship really threatened by our analysis?\footnote{If
asymptotically flat spacetimes are part of the hypothesis of Cosmic
Censorship, as is often the case, then of course no demonstration in
global anti-de Sitter space is relevant.  We prefer a broader
interpretation of Cosmic Censorship---loosely speaking, that no
observer who follows a timelike trajectory which never runs into
singularities can receive signals from a singularity.}  It is too
early to say.  Using the heuristic method of calculating the most
unstable eigenvector of the Hessian of the entropy function, we have
argued that adiabatic evolution of unstable black holes does not lead
to nakedly singular solutions.  However this does not bear directly on
the question of whether the horizon should split as it is assumed to
do in the evolution of unstable black branes.  We leave open many
questions as to the eventual fate of unstable black holes in AdS:
Might they settle down to new non-uniform stationary solutions?  Do
regions of strong curvature form?  Does the horizon split?  Does AdS
itself fragment through a classical process?  We leave these issues
for future work.

\section*{Acknowledgements}

We thank C.~Callan and A.~Chamblin for useful discussions, and
H.~Reall, T.~Pres\-tidge, and G.~Horowitz for other enlightening
communications.  This work was supported in part by DOE
grant~DE-FG02-91ER40671, and by a DOE Outstanding Junior Investigator
award.  S.S.G.\ thanks the Aspen Center for Physics for hospitality
during the early phases of the project.

\bibliography{tachyon}

\begingroup\raggedright\begin{thebibliography}{10}

\bibitem{glOne}
R.~Gregory and R.~Laflamme, ``Black strings and p-branes are unstable,'' {\em
  Phys. Rev. Lett.} {\bf 70} (1993) 2837,
  \href{http://xxx.lanl.gov/abs/hep-th/9301052}{{\tt hep-th/9301052}}.

\bibitem{glTwo}
R.~Gregory and R.~Laflamme, ``The Instability of charged black strings and
  p-branes,'' {\em Nucl. Phys.} {\bf B428} (1994) 399--434,
  \href{http://xxx.lanl.gov/abs/hep-th/9404071}{{\tt hep-th/9404071}}.

\bibitem{witHolTwo}
E.~Witten, ``Anti-de Sitter space, thermal phase transition, and confinement in
  gauge theories,'' {\em Adv. Theor. Math. Phys.} {\bf 2} (1998) 505,
  \href{http://xxx.lanl.gov/abs/hep-th/9803131}{{\tt hep-th/9803131}}.

\bibitem{Ooguri}
C.~Csaki, H.~Ooguri, Y.~Oz, and J.~Terning, ``Glueball mass spectrum from
  supergravity,'' {\em JHEP} {\bf 01} (1999) 017,
  \href{http://xxx.lanl.gov/abs/hep-th/9806021}{{\tt hep-th/9806021}}.

\bibitem{MAGOO}
O.~Aharony, S.~S. Gubser, J.~Maldacena, H.~Ooguri, and Y.~Oz, ``Large N field
  theories, string theory and gravity,'' {\em Phys. Rept.} {\bf 323} (2000)
  183, \href{http://xxx.lanl.gov/abs/hep-th/9905111}{{\tt hep-th/9905111}}.

\bibitem{juanAdS}
J.~Maldacena, ``The large N limit of superconformal field theories and
  supergravity,'' {\em Adv. Theor. Math. Phys.} {\bf 2} (1998) 231--252,
  \href{http://xxx.lanl.gov/abs/hep-th/9711200}{{\tt hep-th/9711200}}.

\bibitem{gkPol}
S.~S. Gubser, I.~R. Klebanov, and A.~M. Polyakov, ``Gauge theory correlators
  from non-critical string theory,'' {\em Phys. Lett.} {\bf B428} (1998) 105,
  \href{http://xxx.lanl.gov/abs/hep-th/9802109}{{\tt hep-th/9802109}}.

\bibitem{witHolOne}
E.~Witten, ``Anti-de Sitter space and holography,'' {\em Adv. Theor. Math.
  Phys.} {\bf 2} (1998) 253--291,
  \href{http://xxx.lanl.gov/abs/hep-th/9802150}{{\tt hep-th/9802150}}.

\bibitem{bkl}
V.~Balasubramanian, P.~Kraus, and A.~Lawrence, ``Bulk vs. boundary dynamics in
  anti-de Sitter spacetime,'' {\em Phys. Rev.} {\bf D59} (1999) 046003,
  \href{http://xxx.lanl.gov/abs/hep-th/9805171}{{\tt hep-th/9805171}}.

\bibitem{gspin}
S.~S. Gubser, ``Thermodynamics of spinning D3-branes,'' {\em Nucl. Phys.} {\bf
  B551} (1999) 667, \href{http://xxx.lanl.gov/abs/hep-th/9810225}{{\tt
  hep-th/9810225}}.

\bibitem{cgOne}
M.~Cvetic and S.~S. Gubser, ``Phases of R-charged black holes, spinning branes
  and strongly coupled gauge theories,'' {\em JHEP} {\bf 04} (1999) 024,
  \href{http://xxx.lanl.gov/abs/hep-th/9902195}{{\tt hep-th/9902195}}.

\bibitem{cgTwo}
M.~Cvetic and S.~S. Gubser, ``Thermodynamic stability and phases of general
  spinning branes,'' {\em JHEP} {\bf 07} (1999) 010,
  \href{http://xxx.lanl.gov/abs/hep-th/9903132}{{\tt hep-th/9903132}}.

\bibitem{Cai}
R.-G. Cai and K.-S. Soh, ``Localization instability in the rotating D-branes,''
  {\em JHEP} {\bf 05} (1999) 025,
  \href{http://xxx.lanl.gov/abs/hep-th/9903023}{{\tt hep-th/9903023}}.

\bibitem{HarmarkOne}
T.~Harmark and N.~A. Obers, ``Thermodynamics of spinning branes and their dual
  field theories,'' {\em JHEP} {\bf 01} (2000) 008,
  \href{http://xxx.lanl.gov/abs/hep-th/9910036}{{\tt hep-th/9910036}}.

\bibitem{cejmOne}
A.~Chamblin, R.~Emparan, C.~V. Johnson, and R.~C. Myers, ``Charged AdS black
  holes and catastrophic holography,'' {\em Phys. Rev.} {\bf D60} (1999)
  064018, \href{http://xxx.lanl.gov/abs/hep-th/9902170}{{\tt hep-th/9902170}}.

\bibitem{cejmTwo}
A.~Chamblin, R.~Emparan, C.~V. Johnson, and R.~C. Myers, ``Holography,
  thermodynamics and fluctuations of charged AdS black holes,'' {\em Phys.
  Rev.} {\bf D60} (1999) 104026,
  \href{http://xxx.lanl.gov/abs/hep-th/9904197}{{\tt hep-th/9904197}}.

\bibitem{gMitra}
S.~S. Gubser and I.~Mitra, ``Instability of charged black holes in anti-de
  Sitter space,'' \href{http://xxx.lanl.gov/abs/hep-th/0009126}{{\tt
  hep-th/0009126}}.

\bibitem{Hawking}
S.~W. Hawking and H.~S. Reall, ``Charged and rotating AdS black holes and their
  CFT duals,'' {\em Phys. Rev.} {\bf D61} (2000) 024014,
  \href{http://xxx.lanl.gov/abs/hep-th/9908109}{{\tt hep-th/9908109}}.

\bibitem{HawkingStrings}
S. Hawking, ``Stability in ADS and Phase Transitions,'' talk at {\it Strings
  '99}, {\tt
  http://strings99.aei-potsdam.mpg.de/cgi-bin/viewit.cgi?speaker=Haw\-king}.

\bibitem{bdhm}
T.~Banks, M.~R. Douglas, G.~T. Horowitz, and E.~Martinec, ``AdS dynamics from
  conformal field theory,'' \href{http://xxx.lanl.gov/abs/hep-th/9808016}{{\tt
  hep-th/9808016}}.

\bibitem{HawkingPage}
S.~W. Hawking and D.~N. Page, ``Thermodynamics of black holes in anti-de Sitter
  space,'' {\em Commun. Math. Phys.} {\bf 87} (1983) 577.

\bibitem{Perry}
D.~J. Gross, M.~J. Perry, and L.~G. Yaffe, ``Instability of flat space at
  finite temperature,'' {\em Phys. Rev.} {\bf D25} (1982) 330--355.

\bibitem{PrestidgeOne}
T.~Prestidge, ``Dynamic and thermodynamic stability and negative modes in
  Schwarzschild-anti-de Sitter,'' {\em Phys. Rev.} {\bf D61} (2000) 084002,
  \href{http://xxx.lanl.gov/abs/hep-th/9907163}{{\tt hep-th/9907163}}.

\bibitem{PrestidgeTwo}
T.~Prestidge, ``Making \$ense of the information loss paradox,''. D.Phil.
  thesis, Cambridge University, January 2000.

\bibitem{PecaOne}
C.~S. Peca and J.~P.~S. Lemos, ``Thermodynamics of Reissner-Nordstroem anti-de
  Sitter black holes in the grand canonical ensemble,'' {\em Phys. Rev.} {\bf
  D59} (1999) 124007, \href{http://xxx.lanl.gov/abs/gr-qc/9805004}{{\tt
  gr-qc/9805004}}.

\bibitem{PecaTwo}
C.~S. Peca and J.~P.~S. Lemos, ``Thermodynamics of toroidal black holes,'' {\em
  J. Math. Phys.} {\bf 41} (2000) 4783--4789,
  \href{http://xxx.lanl.gov/abs/gr-qc/9809029}{{\tt gr-qc/9809029}}.

\bibitem{Price}
R.~H. Price, ``Nonspherical perturbations of relativistic gravitational
  collapse. 1. Scalar and gravitational perturbations,'' {\em Phys. Rev.} {\bf
  D5} (1972) 2419--2438.

\bibitem{ghpss}
S.~B. Giddings, J.~A. Harvey, J.~G. Polchinski, S.~H. Shenker, and
  A.~Strominger, ``Hairy black holes in string theory,'' {\em Phys. Rev.} {\bf
  D50} (1994) 6422--6426, \href{http://xxx.lanl.gov/abs/hep-th/9309152}{{\tt
  hep-th/9309152}}.

\bibitem{sahakian}
D. Sahakian, to appear.

\bibitem{br}
M.~Berkooz and M.~Rozali, ``Near Hagedorn dynamics of NS fivebranes, or a new
  universality class of coiled strings,'' {\em JHEP} {\bf 05} (2000) 040,
  \href{http://xxx.lanl.gov/abs/hep-th/0005047}{{\tt hep-th/0005047}}.

\bibitem{ho}
T.~Harmark and N.~A. Obers, ``Hagedorn behaviour of little string theory from
  string corrections to NS5-branes,'' {\em Phys. Lett.} {\bf B485} (2000)
  285--292, \href{http://xxx.lanl.gov/abs/hep-th/0005021}{{\tt
  hep-th/0005021}}.

\bibitem{kkk}
V. Kazakov, I. Kostov, and D. Kutasov, to appear.

\bibitem{deWitOne}
B.~de~Wit and H.~Nicolai, ``N=8 supergravity with local SO(8) X SU(8)
  invariance,'' {\em Phys. Lett.} {\bf B108} (1982) 285.

\bibitem{deWitTwo}
B.~de~Wit and H.~Nicolai, ``N=8 supergravity,'' {\em Nucl. Phys.} {\bf B208}
  (1982) 323.

\bibitem{deWitThree}
B.~de~Wit and H.~Nicolai, ``The consistency of the $S^7$ truncation in d = 11
  supergravity,'' {\em Nucl. Phys.} {\bf B281} (1987) 211.

\bibitem{DuffLiu}
M.~J. Duff and J.~T. Liu, ``Anti-de Sitter black holes in gauged N = 8
  supergravity,'' {\em Nucl. Phys.} {\bf B554} (1999) 237,
  \href{http://xxx.lanl.gov/abs/hep-th/9901149}{{\tt hep-th/9901149}}.

\bibitem{msBrill}
J.~Maldacena, J.~Michelson, and A.~Strominger, ``Anti-de Sitter
  fragmentation,'' {\em JHEP} {\bf 02} (1999) 011,
  \href{http://xxx.lanl.gov/abs/hep-th/9812073}{{\tt hep-th/9812073}}.

\bibitem{klt}
P.~Kraus, F.~Larsen, and S.~P. Trivedi, ``The Coulomb branch of gauge theory
  from rotating branes,'' {\em JHEP} {\bf 03} (1999) 003,
  \href{http://xxx.lanl.gov/abs/hep-th/9811120}{{\tt hep-th/9811120}}.

\bibitem{fgpwTwo}
D.~Z. Freedman, S.~S. Gubser, K.~Pilch, and N.~P. Warner, ``Continuous
  distributions of D3-branes and gauged supergravity,'' {\em JHEP} {\bf 07}
  (2000) 038, \href{http://xxx.lanl.gov/abs/hep-th/9906194}{{\tt
  hep-th/9906194}}.

\bibitem{cglp}
M.~Cvetic, S.~S. Gubser, H.~Lu, and C.~N. Pope, ``Symmetric potentials of
  gauged supergravities in diverse dimensions and Coulomb branch of gauge
  theories,'' {\em Phys. Rev.} {\bf D62} (2000) 086003,
  \href{http://xxx.lanl.gov/abs/hep-th/9909121}{{\tt hep-th/9909121}}.

\bibitem{cort}
C.~Csaki, Y.~Oz, J.~Russo, and J.~Terning, ``Large N {QCD} from rotating
  branes,'' {\em Phys. Rev.} {\bf D59} (1999) 065012,
  \href{http://xxx.lanl.gov/abs/hep-th/9810186}{{\tt hep-th/9810186}}.

\bibitem{NP}
E.~Newman and R.~Penrose, ``An Approach to gravitational radiation by a method
  of spin coefficients,'' {\em J. Math. Phys.} {\bf 3} (1962) 566--578.

\bibitem{Wald}
R.~M. Wald, {\em General Relativity}.
\newblock Chicago University Press, Chicago, 1984.

\bibitem{gdyad}
S.~S. Gubser, ``Absorption of photons and fermions by black holes in four
  dimensions,'' {\em Phys. Rev.} {\bf D56} (1997) 7854--7868,
  \href{http://xxx.lanl.gov/abs/hep-th/9706100}{{\tt hep-th/9706100}}.

\bibitem{Teukolsky}
S.~A. Teukolsky, ``Perturbations of a rotating black hole. 1. Fundamental
  equations for gravitational, electromagnetic, and neutrino field
  perturbations,'' {\em Astrophys. J.} {\bf 185} (1973) 635--647.

\bibitem{sv}
A.~Strominger and C.~Vafa, ``Microscopic Origin of the Bekenstein-Hawking
  Entropy,'' {\em Phys. Lett.} {\bf B379} (1996) 99--104,
  \href{http://xxx.lanl.gov/abs/hep-th/9601029}{{\tt hep-th/9601029}}.

\bibitem{PeetTASI}
A.~W. Peet, ``TASI lectures on black holes in string theory,''
  \href{http://xxx.lanl.gov/abs/hep-th/0008241}{{\tt hep-th/0008241}}.

\bibitem{gkPeet}
S.~S. Gubser, I.~R. Klebanov, and A.~W. Peet, ``Entropy and Temperature of
  Black 3-Branes,'' {\em Phys. Rev.} {\bf D54} (1996) 3915--3919,
  \href{http://xxx.lanl.gov/abs/hep-th/9602135}{{\tt hep-th/9602135}}.

\bibitem{AndyUnp}
A. Strominger, unpublished, 1996.

\bibitem{hhOne}
G.~T. Horowitz and V.~E. Hubeny, ``Quasinormal modes of AdS black holes and the
  approach to thermal equilibrium,'' {\em Phys. Rev.} {\bf D62} (2000) 024027,
  \href{http://xxx.lanl.gov/abs/hep-th/9909056}{{\tt hep-th/9909056}}.

\bibitem{hhTwo}
G.~T. Horowitz and V.~E. Hubeny, ``CFT description of small objects in AdS,''
  {\em JHEP} {\bf 10} (2000) 027,
  \href{http://xxx.lanl.gov/abs/hep-th/0009051}{{\tt hep-th/0009051}}.

\end{thebibliography}\endgroup
\bibliographystyle{ssg}

\vfil

\end{document}